\documentstyle[twocolumn,aps,prl,floats]{revtex}
\begin{document}
% the following two lines are to get a wide abstract
\twocolumn[\hsize\textwidth\columnwidth\hsize\csname
@twocolumnfalse\endcsname
\title{\bf Controlling domain patterns far from equilibrium}
\draft
\author{A. Hagberg}
\address{Center for Nonlinear Studies, Theoretical Division,
Los Alamos National Laboratory, Los Alamos, NM 87545}
\author{E. Meron}
\address{The Jacob Blaustein Institute for Desert Research and
The Physics Department, Ben-Gurion University\\
Sede Boker Campus 84990, Israel}
\author{I. Rubinstein and B. Zaltzman}
\address{The Jacob Blaustein Institute for Desert Research and
The Mathematics Department, Ben-Gurion University\\
Sede Boker Campus 84990, Israel}

\date{Received 29 August 1995 \\
Phys. Rev. Lett. {\bf 76}, 427 (1996)}
\maketitle
\vspace{-0.2in}
\begin{abstract}

A high degree of control over the structure and dynamics of domain
patterns in nonequilibrium systems can be achieved by applying
nonuniform external fields near parity breaking front bifurcations.
An external field with a linear spatial profile stabilizes a
propagating front at a fixed position or induces oscillations with
frequency that scales like the square root of the field gradient.
Nonmonotonic profiles produce a variety of patterns with
controllable wavelengths, domain sizes, and frequencies and phases
of oscillations.

\end{abstract}
\vspace{-0.3in}
\vskip2pc]
\narrowtext

Technological applications of pattern forming systems are largely
unexplored. The few applications that have been pursued, however,
have had enormous technological impacts.  Magnetic domain patterns
in memory devices provide an excellent example [1].  Intensive
research effort has been devoted recently to dissipative systems
held far from thermal equilibrium [2]. Unlike magnetic materials,
such systems are nongradient in general and their asymptotic
behaviors need not be stationary; a variety of dynamical behaviors
can be realized, including planar and circular traveling waves,
rotating spiral waves, breathing structures and spatiotemporal
chaos. This wealth of behaviors opens up new opportunities for
potential technological applications. Their realizations, however,
depend on the ability to control spatiotemporal patterns by weak
external forces.  Most studies in this direction have focused on
drifting localized structures [3].

In this paper we present a novel way to control domain patterns far
from equilibrium. We consider dissipative systems exhibiting parity
breaking front bifurcations (also referred to as nonequilibrium
Ising-Bloch or NIB transitions [4,5]), in which stationary fronts
lose stability to pairs of counterpropagating fronts.  Examples of
systems exhibiting NIB bifurcations include liquid crystals [6] and
anisotropic ferromagnets [7] subjected to rotating magnetic fields,
chains of coupled electrical oscillators [8], the catalytic CO
oxidation on a platinum surface [9,10], the
ferrocyanide-iodate-sulphite (FIS) reaction [11], and semiconductor
etalons [12]. A prominent feature of these systems is that
transitions between the parity broken states, the left and right
propagating fronts, become feasible as the front bifurcation is
approached.  Indeed, intrinsic disturbances, like front curvature
and front interactions, are sufficient to induce spontaneous
transitions and can lead to complex pattern formation phenomena such
as breathing labyrinths, spot replication [11,13] and spiral
turbulence [14].  It is this dynamical flexibility near NIB
bifurcations that we wish to exploit. By forcing transitions between
the left and right propagating fronts, using spatially dependent
external fields, we propose to obtain a high degree of control on
pattern behavior.

We demonstrate this idea using a forced activator-inhibitor system
of the form
%%%%%%%%%%%%%%%%%%%%%%%%%%%  1  %%%%%%%%%%%%%%%%%%%%%%%%%%%%%%%%%%%%%%%%%
\begin{eqnarray}
u_t&=&\epsilon^{-1}(u-u^3-v)+u_{xx},\nonumber \\
v_t&=&u-a_1v+\delta v_{xx}+h+Jv_x, \eqnum{1}
\end{eqnarray}
where $u$, the activator, and $v$, the inhibitor, are scalar real
fields, and $h$ and $J$ are external fields. With appropriate choice
of $a_1>0$, the system (1) has two linearly stable stationary
uniform solutions, an "up" state $(u_+,v_+)$ and a "down" state
$(u_-,v_-)$, and front solutions connecting these states. The domain
patterns to be considered here consist of one dimensional arrays of
up state regions separated by down state regions.  In the absence of
the external fields, a stationary front solution, stable for
$\epsilon>\epsilon_c(\delta)$, loses stability in a pitchfork
bifurcation to a pair of fronts propagating in opposite directions
at constant speed.  For $\epsilon/\delta\ll 1$ the bifurcation point
is given by $\epsilon_c={9\over 8q^6\delta}$ where $q^2=a_1+1/2$
[14]. Activator inhibitor models have been used to describe some of
the systems mentioned above [8,9,10,12]. In these systems $h$ is
usually a parameter that introduces an asymmetry between the up and
down states. In the context of chemical reactions involving ionic
species $J$ may stand for an electric field [15].

\begin{figure}[t]
\vspace{2.5in}
\includegraphics{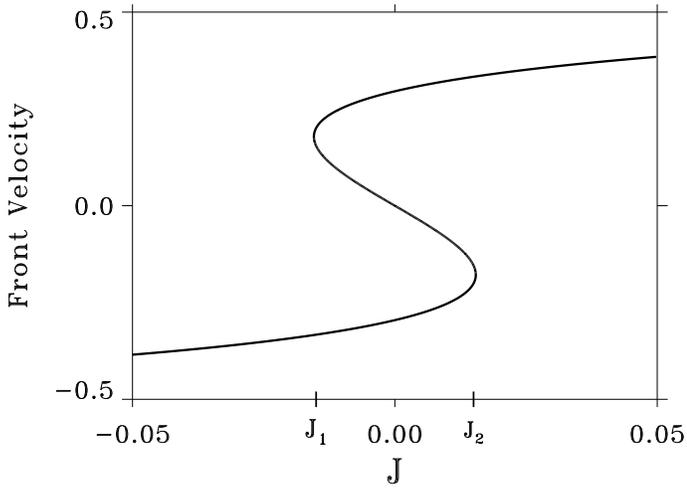}
\caption{
Front velocity vs. external field, $J$, near the nonequilibrium Ising-
Bloch bifurcation.
Parameters: $a_1=3.0,~\epsilon=0.01,~\delta=2.0,~h=0$.\hfil\break\noindent}
\end{figure}
Consider first the effect of a constant external field on the front
velocity.  Away from a front bifurcation the effect of a weak field
is captured well by a linear approximation and therefore the effect of the
field is small.  This is not the case close to a front bifurcation;
the velocity - external field relation becomes multivalued (or
hysteretic) even for a weak field as illustrated in Fig.~1
[8,10,13,16].  
This form is a generic unfolding of a pitchfork
bifurcation and holds for various unfolding forces and parameters
including intrinsic disturbances like curvature [13]. We emphasize
that the termination points, $J_1$ and $J_2$, of the upper and lower
branches lie close to $J=0$. The significance is that {\it weak}
external fields can induce transitions between the two branches, or
reverse the direction of front propagation.
%The upper and lower branches in this figure
%pertain to an up state invading a down state and to a down state invading 
%an up state, respectively. The middle branch represents an unstable front.

The effect of a {\it nonuniform} external field  [17] can be understood 
intuitively in the following way. Consider a constant $h$ and a linear 
profile for $J$: $J=-\alpha x$, where $0<\alpha\ll 1$.
This choice divides space into three regions according to the type 
and number of existing front solutions: (i) $x>x_1=-J_1/\alpha$, where
$J<J_1$ and only a single front corresponding to a down state invading an up 
state exists, (ii) $x<x_2=-J_2/\alpha$, where $J>J_2$ and only a single front 
corresponding to an up state invading a down state exists, and (iii) 
$x_2<x<x_1$, where $J_1<J<J_2$ and both fronts coexist. This profile of $J$ 
results in an {\it oscillating} front, with oscillations roughly spanning 
the interval $x_2<x<x_1$. The front propagation direction is reversed 
during transitions from the upper to the lower 
velocity branch at $J\approx J_1$ and from the lower to the upper branch at 
$J\approx J_2$. Obviously, a variety of pattern behaviors can be induced 
using a nonmonotonic $J$ profile. For example, a single hump profile can 
induce a breathing domain.

We now turn to a quantitative study of front dynamics and relate pattern 
characteristics (e.g. breathing frequency) to control parameters. 
We assume $\epsilon\ll 1$ and distinguish between an inner region, the 
narrow front region, $x\approx x_f$, where $u$ varies 
sharply over a distance of ${\cal O}(\sqrt\epsilon)$, and outer regions,
$x<x_f$ and $x>x_f$, where $u$ varies on the same scale as $v$. 

Consider first the inner region. Expressing (1) in a frame moving with 
the front, $x\to r=x-x_f(t)$, stretching the spatial coordinate according 
to $z=r/\sqrt\epsilon$, and expanding  
$u=u_0+\sqrt\epsilon u_1 + \epsilon u_2 +...$,
$~~v=v_0+\sqrt\epsilon v_1 + \epsilon v_2 +...$, 
we find at order unity the stationary front solution, 
$u_0=-\tanh(z/\sqrt 2),~~~v_0=0$. At order $\sqrt\epsilon$ we find
%%%%%%%%%%%%%%%%%%%%%%%%%%%%%  2  %%%%%%%%%%%%%%%%%%%%%%%%%%%%%%%%%%%%%%%
$${\cal L}  u_1= v_1-\dot x_f u_{0z},\qquad {\cal L}=
\partial_z^2+1-3u_0^2.\eqno(2)$$
Solvability of (2) gives
%%%%%%%%%%%%%%%%%%%%%%%%%%%%%  3  %%%%%%%%%%%%%%%%%%%%%%%%%%%%%%%%%%%   
$$\dot y_f =-{3\over \eta\sqrt 2} v_f(t),\eqno(3)$$
up to corrections of ${\cal O}(\epsilon)$, where 
$y_f=x_f/\sqrt\delta$, $\eta^2=\epsilon\delta$, and $v_f(t)=v(0,t)$ is 
the yet undetermined value of the inhibitor at the front position $r=0$.

A dynamical equation for $v_f$ follows from an analysis of the outer
regions. First we go back to the unstretched 
coordinate system and rescale the spatial coordinate according to 
$y=r/\sqrt\delta$. At order unity we find
%%%%%%%%%%%%%%%%%%%%%%%%%%%%%  4  %%%%%%%%%%%%%%%%%%%%%%%%%%%%%%%%%%%% 
$$v_t-\dot y_fv_y=u_+(v)-a_1v+v_{yy}-\alpha(y+y_f)v_y+h  
\qquad y\le0,\eqno(4)$$
%v_t-\dot y_fv_y&=&u_-(v)-a_1v+v_{yy}-\alpha(y+y_f)v_y+h \nonumber \\
%&&y\ge 0,\eqnum{4}
%\end{eqnarray}
and a similar equation for $y\ge 0$ with $u_+(v)$ replaced by $u_-(v)$.
Here $u_\pm(v)$ are the outer solution branches 
of the cubic equation $u-u^3-v=0$. In ($4$) we assumed the field $h$ is
constant and took a linear profile $J=-\alpha x$.
For $a_1$ sufficiently large we may linearize the branches 
$u_\pm(v)$ around $v=0$, $u_\pm(v)\approx \pm 1-v/2$. Inserting (3) in 
(4) and using the approximate forms for $u_\pm(v)$ we find the free 
boundary problem
%%%%%%%%%%%%%%%%%%%%%%%%%%%%%%  5  %%%%%%%%%%%%%%%%%%%%%%%%%%%%%%%%%%%
\begin{eqnarray}
{\cal M}v&=&+1 -\alpha(y+y_f)v_y+h -{3\over \eta\sqrt 2}v(0,t)v_y
\qquad y\le 0\nonumber \\
{\cal M}v&=&-1 -\alpha(y+y_f)v_y+h -{3\over \eta\sqrt2}v(0,t)v_y
\qquad y\ge 0,\nonumber
\end{eqnarray}
$$v(\pm\infty,t)=v_\mp= \mp
q^{-2}\qquad \bigl[v\bigr]_{y=0}=\bigl[v_y\bigr]_{y=0}=0,\eqno(5)$$ 
where ${\cal M}=\partial_t+q^2-\partial_x^2$,  
the square brackets denote jumps across the free boundary at $y=0$, and
$v(0,t)$ and $y_f(t)$ satisfy (3).

To solve the free boundary problem (5) we assume the system is close to 
the front bifurcation so that the speed, $c$,  of the propagating solutions
in the absence of the external fields is small. We also take the external 
fields to be of order $c^3$: $\alpha=\alpha_0 c^3$, $h=h_0 c^3$. We now 
expand the propagating solutions as a power series in $c$:
%%%%%%%%%%%%%%%%%%%%%%%%%%%%%%%  6  %%%%%%%%%%%%%%%%%%%%%%%%%%%%%%%%%%%%%%%%
$$v(y,t,T)=v^{(0)}(y)+\sum_{n=1}^\infty c^n v^{(n)}(y,t,T),\eqno(6)$$
where $v^{(0)}(y)$, the stationary front solution, is an odd function given
by $v^{(0)}(y)=q^{-2}(e^{-qy}-1)$ for $ y\ge 0$,
%%%%%%%%%%%%%%%%%%%%%%%%%%%%%%%  7  %%%%%%%%%%%%%%%%%%%%%%%%%%%%%%%%%%%%%%
%\begin{eqnarray}
%v^{(0)}(y)=&q^{-2}(1-e^{qy}),&\qquad y\le 0,\nonumber \\
%v^{(0)}(y)=&q^{-2}(e^{-qy}-1),&\qquad y\ge 0,\eqnum{7}
%\end{eqnarray}
and $T=c^2 t$ is a slow time scale
characterizing the nonsteady front motion near the bifurcation.
Expanding $\eta$ as well, $\eta=\eta_c-c^2\eta_1+c^4\eta_2+...,~$
and inserting these expansions  in (5) we find
%%%%%%%%%%%%%%%%%%%%%%%%%%%%%%%  7  %%%%%%%%%%%%%%%%%%%%%%%%%%%%%%%%%%%%%%%
$$v^{(n)}_t+q^2v^{(n)}-v^{(n)}_{yy}=-\rho^{(n)}\qquad n=1,2,3,\eqno(7)$$
where
%%%%%%%%%%%%%%%%%%%%%%%%%%%%%%%  8 %%%%%%%%%%%%%%%%%%%%%%%%%%%%%%%%%%%%%%%
\begin{eqnarray}
\rho^{(1)}&=&{3\over\sqrt2\eta_c}v^{(1)}_{\vert y=0}v^{(0)}_y,  \nonumber \\
\rho^{(2)}&=&{3\over\sqrt2\eta_c}\left[v^{(1)}_{\vert y=0}v^{(1)}_y+
v^{(2)}_{\vert y=0}v^{(0)}_y\right],  \nonumber \\
\rho^{(3)}&=&v^{(1)}_T+{3\eta_1\over\sqrt
2\eta_c^2}v^{(1)}_{\vert y=0}v^{(0)}_y  \nonumber \\
&&+{3\over\sqrt2\eta_c}\left[v^{(1)}_{\vert y=0}v^{(2)}_y  
+v^{(2)}_{\vert y=0}v^{(1)}_y+ v^{(3)}_{\vert y=0}v^{(0)}_y\right] \nonumber \\
&&+\alpha_0(y+y_f)v^{(0)}_y - h_0.
\eqnum{8}
\end{eqnarray}
The solution of (7) with a zero initial condition (only relevant as long
as the long fast time asymptotics is concerned) satisfies the integral 
equation
%%%%%%%%%%%%%%%%%%%%%%%%%%%%%%  9  %%%%%%%%%%%%%%%%%%%%%%%%%%%%%%%%%%%%%%%%%
\begin{eqnarray}
v^{(n)}(y,t,T)&=&-\int_0^t d\tau~{e^{-q^2(t-\tau)}\over 2[\pi(t-\tau)]^{1/2}}
\eqnum{9} \\
&&\times \int_{-\infty}^\infty d\xi ~exp\left(-{(y-\xi)^2\over 4(t-\tau)}\right)
\rho^{(n)}(\xi,\tau,T).
\nonumber
\end{eqnarray}
Recall that $\rho^{(n)}$ contains the
unknown $v^{(n)}$ evaluated at $y=0$. Since the origin of the 
slow time scale is the nonsteady front motion we expect $v^{(n)}_{\vert y=0}$
to become independent of the fast time scale $t$ as $t\to\infty$. 
Substituting $\lim_{t\to\infty}v^{(n)}(0,t,T)=v^{(n)}(0,T)$ in (9) and 
setting $y=0$ we find a sequence of compatibility conditions. The 
first, for $n=1$, is $\eta_c={3\over 2\sqrt 2 q^3}$.
The critical value $\eta_c=\eta(c=0)$ determines the front bifurcation 
point. The compatibility condition for $n=3$ is 
%%%%%%%%%%%%%%%%%%%%%%%%%%%%%%%%%  10  %%%%%%%%%%%%%%%%%%%%%%%%%%%%%%%%%%%
\begin{eqnarray}
v^{(1)}_T(0,T)&=&{\sqrt 2\eta_1\over q\eta_c^2} v^{(1)}(0,T)-{3\over
4\eta_c^2}{v^{(1)}}(0,T)^3 \nonumber \\
&&+{2\over 3q}\alpha_0 y_f+{4\over 3}h_0, 
\eqnum{10}
\end{eqnarray}
where $\eta_1={q\eta_c^2\over 6\sqrt 2\delta}$. Expressing (10) in terms of 
$v_f=v(0,T)
\approx cv^{(1)}(0,T)$ and using $c^2\eta_1\approx\eta_c-\eta$ we find 
%%%%%%%%%%%%%%%%%%%%%%%%%%%%%%%%%  11 %%%%%%%%%%%%%%%%%%%%%%%%%%%%%%%%%%%
\begin{equation}
\dot v_f={\sqrt 2\over q\eta_c^2}(\eta_c-\eta) v_f-{3\over
4\eta_c^2}v_f^3+{2\over 3q}\alpha y_f+{4\over 3}h,\eqnum{11} 
\end{equation}
where $\dot v_f=c^2 {v_f}_T$.

Equations (3) and (11) describe the dynamics of fronts near a NIB
bifurcation, subjected to a constant field $h$ and a linearly space
dependent $J$ field.  In addition to the translational degree of
freedom, $y_f$, the dynamics involve a second degree of freedom, the
value of the inhibitor at the front position, $v_f$, which is responsible 
for 
transitions between the left and right propagating fronts. Without the
field $J$ the system (3) and (11) is decoupled and reproduces the front
bifurcation. The introduction of a {\it space dependent} field
couples the two degrees of freedom and affects the front behavior in
two significant ways: for $\eta>\eta_c$ (and $h\ne 0$) it stabilizes
a propagating front at a fixed position, $y_f=-{2qh\over \alpha}$,
and for $\eta<\eta_c$ it induces oscillations between the
counterpropagating fronts. The frequency of oscillations close to the
Hopf bifurcation at $\eta=\eta_c$ is
%%%%%%%%%%%%%%%%%%%%%%%%%%%%%%%%  12  %%%%%%%%%%%%%%%%%%%%%%%%%%%%%%%%%
$$\omega={2\over\sqrt 3}q\sqrt\alpha.\eqno(12)$$

\begin{figure}[thb]
\vspace{4.7in}
\includegraphics{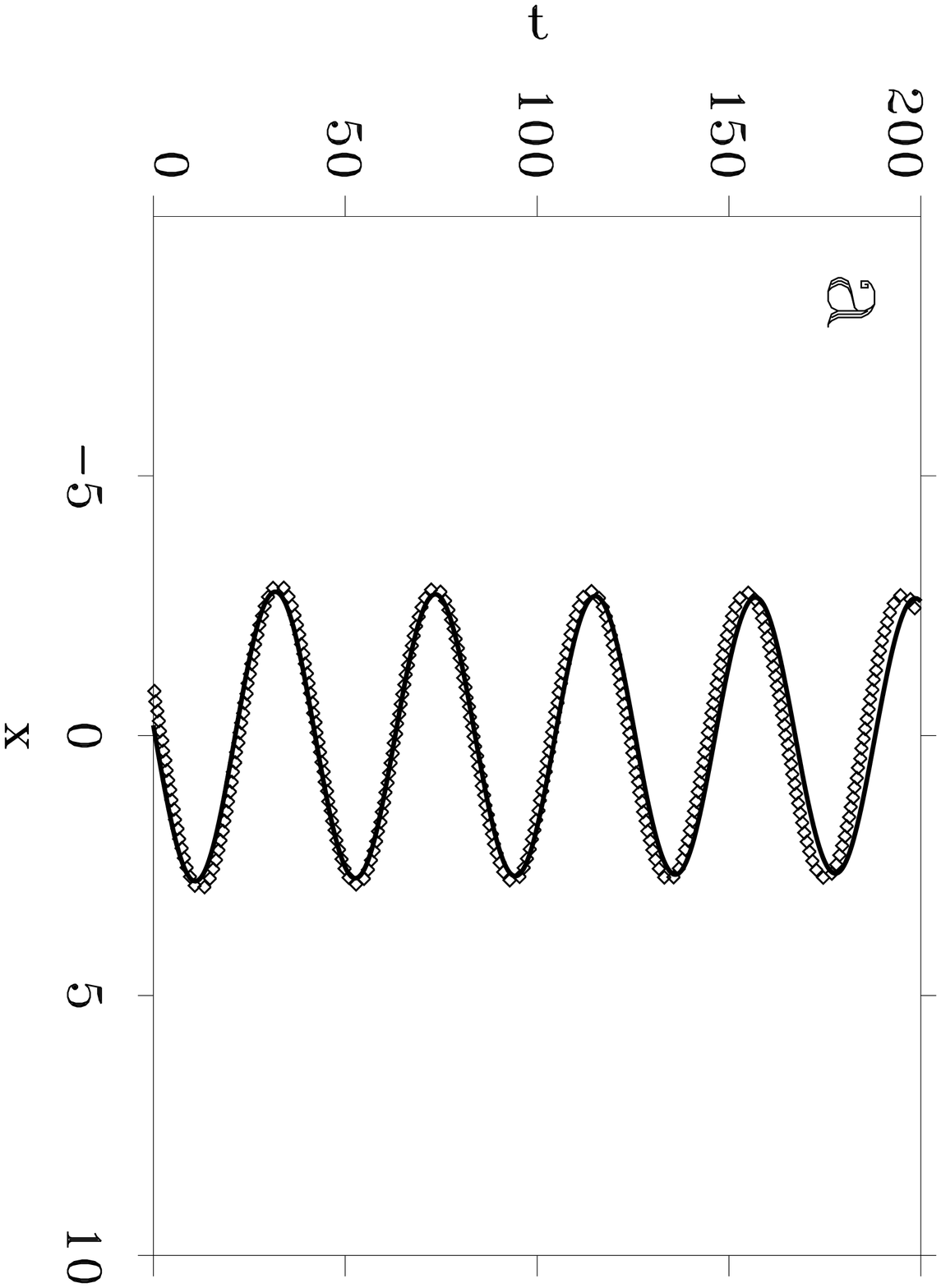}
\includegraphics{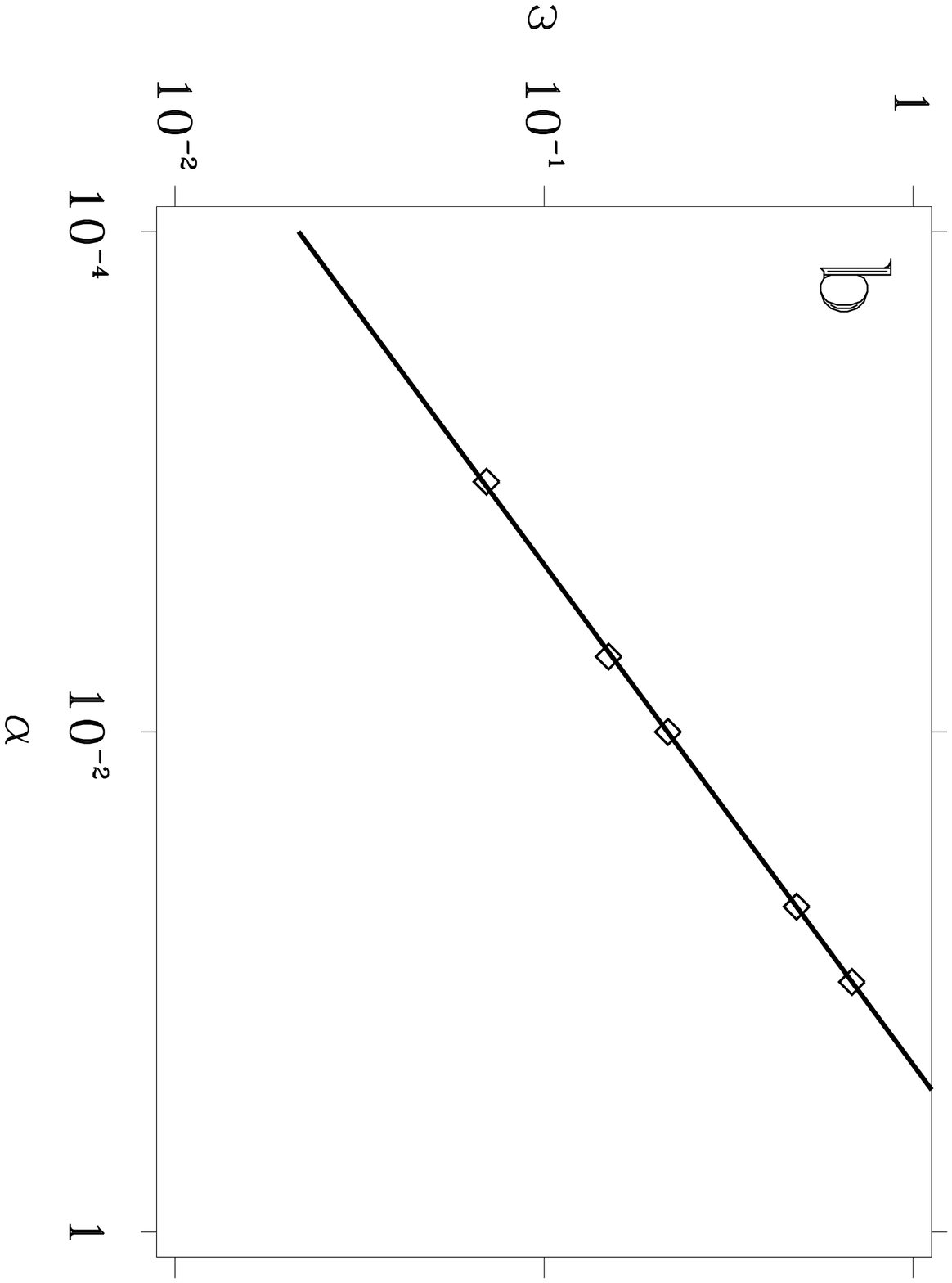}
\caption{
(a) Front position, $x_f$, vs. time for an oscillating front.  The
thin line represents the solution to equations (3) and (11) and the diamonds
are from the numerical solution of equations (1). 
(b) A log-log plot of the oscillation frequency, $\omega$, 
vs. the external field
gradient, $\alpha$. The solid line is the relation of equation (12) and
the diamonds represent numerical solutions of equations (1). 
Parameters: $a_1=3.0,~\epsilon=0.01,~\delta=2.77,~h=0$ and $\alpha=0.005$
in (a).}
\end{figure}
To test the validity of equations (3) and (11) we numerically integrated the 
original 
system (1) and compared oscillating front solutions of (1) with those of (3) 
and (11). The
agreement as Fig. 2$a$ shows is very good. In Fig. 2$b$ we plotted the 
frequency of front oscillations vs. the field gradient according to (12)
and as obtained from (1). 
Again, the agreement is excellent, and 
remains good even for $c$ of order unity.

\begin{figure}[t]
\vspace{2.5in}
\includegraphics{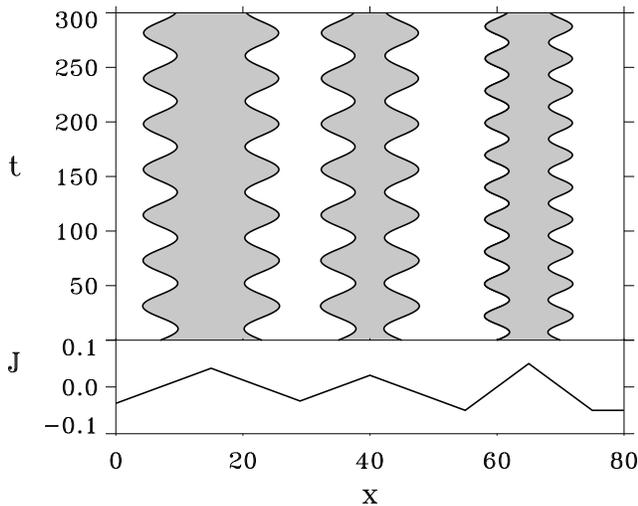}
\caption{
%The average size and frequency of oscillating domains can be controlled 
%by varying the spatial profile of the external field. 
The oscillating domain on the left has the same 
frequency as middle domain but the wider separation 
between the $J=0$ points produces a larger domain size. 
The domain on the right has the same average width as 
the middle domain but the larger external field 
gradient produces higher frequency. 
Parameters: $a_1=3.0,~\epsilon=0.01,~\delta=2.77,~h=0$.
}
\end{figure}
These results suggest various ways to control domain
patterns. \begin{figure}[h]
\vspace{2.6in}
\includegraphics{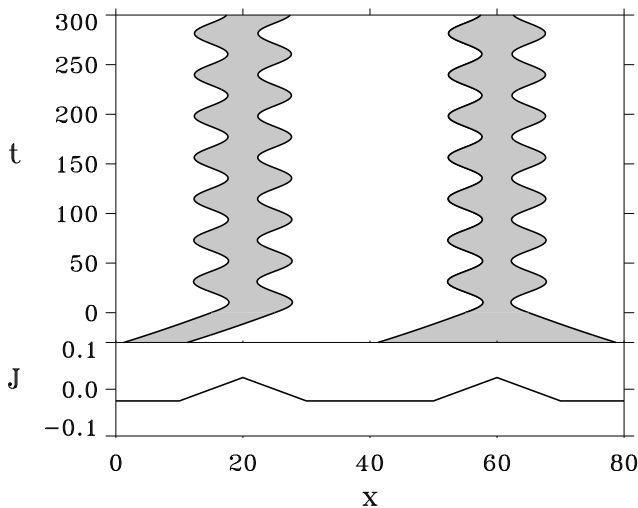}
\caption{
The phase of oscillation is determined by the 
$v_f$ values of the fronts that bound a domain. Opposite signs of 
$v_f$ give rise to a traveling domain that oscillates back and forth 
(left domain).  Equal signs of $v_f$ yield a breathing domain 
(right domain). 
%The $v_f$ values were selected by starting with a traveling
%domain and two approaching fronts in the absence of $J$
%and then switching $J(x)$ on at $t=0$.
Parameters: $a_1=3.0,~\epsilon=0.01,~\delta=2.77,~h=0$.
}
\end{figure}
Choosing a periodic $J(x)$ profile, for example, allows
the creation of a periodic pattern of stationary ($\eta>\eta_c$) or
oscillating ($\eta<\eta_c$) domains. The period of the pattern, the
average width of the domains, and the frequency and relative phase
of oscillations are easily controllable. Figs. 3 and 4 show the
effect of a nonuniform triangular profile of $J$ for
$\eta<\eta_c$. 
In the absence of $J$ the system supports traveling
domain patterns. Switching on $J(x)$ gives rise to patterns of
oscillating domains. Regions of $J(x)$ where the gradient
$\alpha(x)=\vert J^\prime(x)\vert$ is steeper yield higher
oscillation frequencies in accord with (12), while wider $J$
triangles yield wider domains (Fig. 3). The relative phase of
oscillation is controlled by the values of $v_f$ for the two fronts
that bound a domain. Choosing the same sign for $v_f$ gives rise to
breathing dynamics whereas opposite signs yield back and forth
oscillations (Fig. 4). For $\eta>\eta_c$ arbitrary stationary domain
patterns can be formed with appropriate $J(x)$ profiles; the only
restriction is the requirement of a minimum domain size to guarantee
the dominance of $J$ over front interactions. Similar results are
obtained with a nonuniform $h$ field and constant $J$.  
We expect the main ideas
presented here to apply to other models exhibiting NIB bifurcations, 
such as the forced complex Ginzburg-Landau equation [4].

\end{document}